\documentclass[10pt,sigconf,letterpaper]{acmart}
\renewcommand\footnotetextcopyrightpermission[1]{} 
\setcopyright{none}
\usepackage{ifthen}
\usepackage{soul}
\usepackage{csquotes}
\usepackage{amsmath}
\usepackage[inline]{enumitem}
\usepackage{algorithm}
\usepackage{algpseudocode}

\usepackage{algpseudocode}
\usepackage{etoolbox, xstring}
\usepackage{xspace}
\DeclareListParser{\doslashlist}{/}
\newcounter{ndnNameComponentCounter}%
\newcommand{\name}[1]{{%
		\setcounter{ndnNameComponentCounter}{0}%
		\renewcommand{\do}[1]{{%
				\ifnumgreater{\value{ndnNameComponentCounter}}{0}{\allowbreak/}{}%
				\ifnumodd{\value{ndnNameComponentCounter}}{}{}%
				\detokenize{##1}}%
			\stepcounter{ndnNameComponentCounter}}%
		``{\fontfamily{cmtt}\small\selectfont\IfBeginWith{#1}{/}{/}{}\doslashlist{#1}}''%
}}

\newboolean{showcomments}
\setboolean{showcomments}{true}
\ifthenelse{\boolean{showcomments}}
{ \newcommand{\mynote}[3]{
    \protect\fbox{\bfseries\sffamily\scriptsize#1}
    {\small$\blacktriangleright$\textsf{\emph{\color{#3}{#2}}}$\blacktriangleleft$}}}
{ \newcommand{\mynote}[3]{}}

\usepackage{xcolor}

\newcommand{\eg}{\textit{e.g.,}\@\xspace}
\newcommand{\ie}{\textit{i.e.,}\@\xspace}

\definecolor{verylightgray}{gray}{0.8}

\usepackage[normalem]{ulem}

\author{Tianyuan Yu}
\affiliation{%
  \institution{UCLA}
  \city{Los Angeles}
  \state{CA}
  \country{USA}
}

\author{Lan Wang}
\affiliation{%
  \institution{University of Memphis}
  \city{Memphis}
  \state{TN}
  \country{USA}
}

\author{Beichuan Zhang}
\affiliation{%
  \institution{University of Arizona}
  \city{Tucson}
  \state{AZ}
  \country{USA}
}

\author{Lixia Zhang}
\affiliation{%
  \institution{UCLA}
  \city{Los Angeles}
  \state{CA}
  \country{USA}
}

\begin{document}

\title{From Map-and-Encap to BIER:\\
Observations on Network Routing Scalability}

\begin{abstract}
The TCP/IP protocol stack uses IP addresses for two distinct roles: identifying hosts and locating their attachment points in the network topology. This dual purpose creates a fundamental tension that has led to routing and forwarding scalability challenges throughout the history of the Internet in unicast packet delivery and, more notably, in multicast delivery.

This paper reviews the evolution of routing scalability solutions over the years and makes four observations. First, map-and-encap is a recurring architectural solution shared by all scalable unicast and multicast delivery methods, developed independently across different problem contexts. Second, a new solution tends to succeed when it can bring immediate local gains to early adopters without requiring coordination across administrative domains. 
Third, network routing and forwarding designs that depend on external factors, such as the number of distinct end sites or even application-specific deliveries, inherently preclude an upper bound on their scalability.
Fourth, today's inter-domain routing protocol, BGP, lacks a topological abstraction equivalent to an egress router within a routing domain, thereby inherently preventing a map-and-encap solution for scalability. 
These observations offer insights into the design of future scalable routing system architectures.
\end{abstract}

\maketitle

\section{Introduction}
\label{sec:intro}

The Internet was originally designed to deliver packets between hosts, using IP addresses as the universal identifiers for packet destinations~\cite{clark88}. 
Each packet carries its destination address, and each router maintains a forwarding table indexed by destination IP address prefixes. 
This design is simple, robust, and has enabled the Internet to grow from a small experimental network into a global network connecting billions of devices.

By the original design, IP addresses serve \emph{two distinct roles} simultaneously:  as \emph{topological locators} indicating an end site's position in the network topology, and as \emph{endpoint identifiers} that site operators embed in long-term configurations, contracts, and access control lists.
As the Internet expanded globally, this dual role created a fundamental tension. 
Network providers desire end sites to use provider-assigned address blocks, which can be aggregated into compact routing entries to scale the routing system.
But when an end site changes providers, its address block must also change to reflect its new topological location, undermining its utility as a stable identifier and forcing adjustments to the site's internal network management.
Conversely, if an end site acquires a provider-independent address block, that block must be advertised globally, which increases the routing table size and reduces aggregation efficiency. 
In other words, \emph{the needs for identification and location pull in opposite directions}. This ongoing tension has driven routing scalability challenges since the 1990s, and a similar but more pronounced conflict underlies the scalability problems associated with multicast delivery.

This paper reviews the evolution of routing scalability solutions, from the early recognition of the Identifier/Locator conflict in unicast and the Map-and-Encap response~\cite{rfc1955}, to the scaling challenges in multicast and the Bit Index Explicit Replication (BIER)~\cite{rfc8279} as a more promising solution direction.
In this review, we make four technical observations.
First, map-and-encap is a recurring solution shared by all scalable unicast and multicast delivery methods.
Second, a new solution tends to succeed when it can bring immediate local gains to early adopters without requiring coordination across administrative domains, while solutions requiring global coordination before anyone sees local benefit face nearly insurmountable deployment barriers. 
Third, network routing-and-forwarding designs should scale with the network topology, a quantity that can be engineered and controlled, rather than with external factors that are outside the network's control. 
Fourth, today's inter-domain routing design lacks a topological abstraction similar to an egress router in a single domain, which explains why no global map-and-encap solution has been successfully designed or deployed.

\section{Unicast Routing Scalability}
\label{sec:unicast}

\subsection{The Identifier/Locator Conflict}
Routing scalability has long been recognized as an open challenge. As the Internet expanded and commercialized, enterprises, institutions, and end users all desired address blocks that were independent of their current provider. This would enable them to switch providers without renumbering and to multi-home with multiple providers for both resiliency and traffic engineering.
However, provider-independent addresses cannot be aggregated by the routing system. The result is the unstoppable growth of the Default-Free Zone (DFZ) routing table, which has increased from tens of thousands of entries in the early 1990s to over one million entries today.
The growth is driven not by the number of routers or links, but by the number of distinct end-site address blocks --- a quantity that increases with administrative and commercial decisions rather than network topology.

This conflict was captured in \emph{Rekhter's Law} in 2007:
\begin{displayquote}
Addressing can follow topology, or topology can follow addressing. Choose one~\cite{fuller2007}.
\end{displayquote}
It states that the dual role of IP addresses as both identifiers and locators cannot be reconciled. A 2007 IAB workshop on routing and addressing~\cite{rfc4984} identified this conflict as the root cause of routing table growth, pointed to map-and-encap as the most promising architectural solution, and directly influenced the development of LISP~\cite{rfc9300}.

\subsection{Map-and-Encap: a Principled Response}
\begin{figure}[h]
    \centering
    \includegraphics[width=0.48\textwidth]{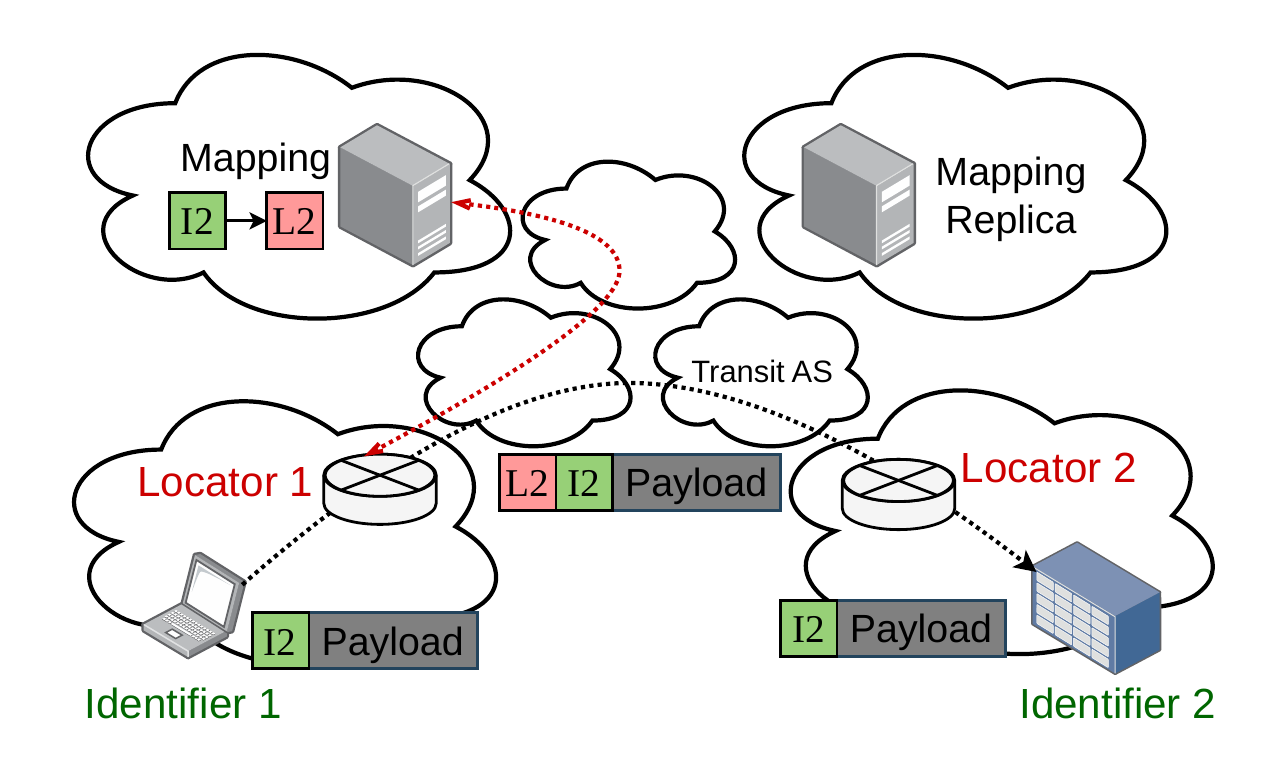}
    \caption{The map-and-encap approach to scale routing. Logical end-site \emph{identifiers} are mapped to topological \emph{locators} at the network boundary, enabling core routers to maintain state for aggregatable provider addresses only.}
    \label{fig:map}
\end{figure}
A principled resolution to the Identifier/Locator conflict is dubbed map-and-encap, proposed in 1992 and formally documented by RFC~1955~\cite{rfc1955} in~1996.
The basic idea is straightforward: separate the two roles of IP addresses into two distinct namespaces.
The end sites retain their addresses as logical identifiers, which are provider-independent.
The addresses assigned to the provider networks are topological locators.
They are aggregatable addresses that reflect their position in the network topology.
At the network boundary, a mapping system translates logical identifiers to topological locators.
RFC~1955 proposed using DNS to provide this mapping service: the ingress router performs a DNS lookup on the destination's logical identifier to obtain the corresponding topological locator, then encapsulates the packet with an outer header carrying that locator as the destination.
Core routers forward based on the outer (topological) header without awareness of the inner logical identifier.

Figure~\ref{fig:map} illustrates a map-and-encap implementation.
Under map-and-encap, the core routing table contains only topological locators: provider addresses, hierarchically assigned, and aggregatable.
Logical identifiers for the end site do not appear in the core routing table.
The forwarding state in the core is bounded by the number of provider networks --- a topological quantity 
--- rather than by the number of end sites, which grows without known upper bounds.

LISP~\cite{rfc9300} is the most fully specified realization of map-and-encap.
It introduces Endpoint Identifiers (EIDs) as logical end-user addresses, and Routing Locators (RLOCs) as topological provider addresses.
There is a separate Delegated Database Tree (DDT) that translates EIDs to RLOCs at the network boundary (see Section~\ref{sec:mapping} for more details about this mapping system).
Ingress tunnel routers (ITRs) encapsulate packets with an outer IP header that carries the RLOC destination; egress tunnel routers (ETRs) decapsulate and deliver to the EID.
The core routers forward based on RLOCs only.

\subsection{Map-and-Encap: The Reality}
\begin{figure}[hbtp]
    \centering    \includegraphics[width=0.48\textwidth]{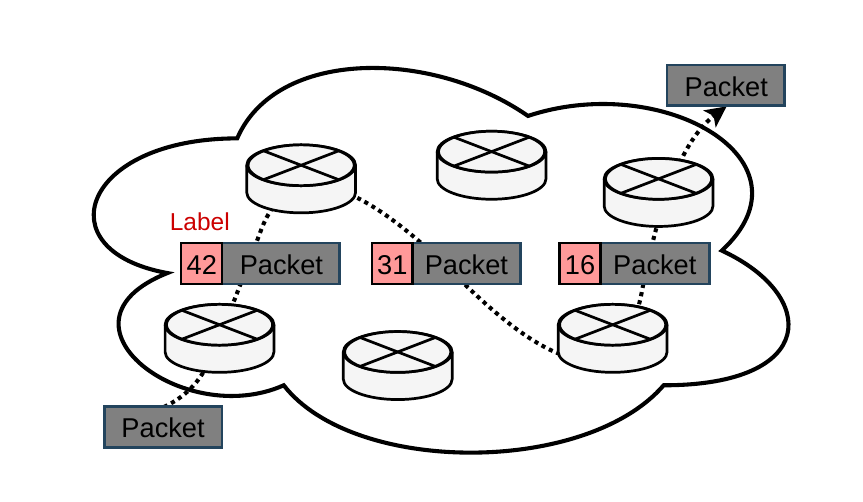}
    \caption{An example scenario of MPLS, which replaced prefix lookups by hop-by-hop label-switched paths.}
    \label{fig:mpls}
\end{figure}

While LISP was first specified in experimental RFCs in 2013~\cite{rfc6830, rfc6833} and later standardized in 2022~\cite{rfc9300, rfc9301, rfc9303}, its deployment in the public Internet has remained limited. 
The industry has largely continued to route user prefixes in the global routing system, supported by steady advances in router hardware (larger TCAM memories, faster lookup engines) and routing algorithms.  
Small ISPs manage table size through default routes. 
Large ISPs avoid the full DFZ forwarding tables \emph{internally} by running Multiprotocol Label Switching (MPLS)~\cite{rfc3031}, as illustrated in Figure~\ref{fig:mpls}: when a packet enters an AS, the ingress router classifies it into a Forwarding Equivalence Class (FEC), selects a corresponding Label-Switched Path (LSP) toward the appropriate egress, and pushes an MPLS label onto the packet.
Core routers then forward packets based on labels rather than performing IP prefix lookups.

In architectural terms, MPLS can be viewed as a form of \emph{map-and-encap within a single administrative domain.}
At ingress, packets are mapped to FECs associated with LSPs that terminate at specific egress routers, with forwarding state encoded in the label stack. However, unlike explicit map-and-encap, MPLS does not need a separate mapping system: the association between FECs and egress points is derived from existing routing protocols (\ie eBGP/iBGP for inter-domain reachability and IGP for intra-domain topology/routing) and distributed via label distribution mechanisms. Additionally, MPLS encapsulation (label stacking) is typically confined within a single provider domain.

A recurring reality in network engineering is that existing deployments resist architectural changes when incremental adaptations remain viable.
Deploying LISP at a global scale requires 
(1) a new mapping system to coordinate Identifier-to-Locator translation at Internet scale, and
(2) modification to border routers across administrative domains. 
Not only do these two steps incur significant costs, but first movers gain little benefit before the critical mass is reached. 

In contrast, advances in hardware and algorithms require only equipment upgrades, and individual adopters can see immediate results. The engineering tradeoff favored the incremental path, and the industry followed it. As a result, intra-domain MPLS has seen widespread deployment, partly because it leverages existing routing protocols and, more importantly, because it makes local decisions under the control of a single operator and provides immediate local benefits (preventing core routers from performing global IP prefix lookups).
In general, a new solution tends to succeed when it can deliver immediate local gains without needing coordination across administrative domains.


\begin{quote}
\noindent\textbf{Observation 1: a new routing solution that demands global coordination faces a nearly insurmountable deployment barrier. MPLS succeeded because it is under a single operator's control and provides immediate local gains. LISP has seen limited deployment because it requires establishing a new global mapping system and cross-domain coordination before early adopters see benefits.}
\end{quote}

The Identifier/Locator conflict in unicast, therefore, remains unresolved at the global scale. Individual operators manage their costs through well-engineered local solutions --- default routes for small ISPs, MPLS for large ones --- but no architectural scalable solution has been deployed globally.

\section{Multicast Routing Scalability}
\label{sec:multicast}
\begin{figure*}[h]
    \centering
    \includegraphics[width=\linewidth]{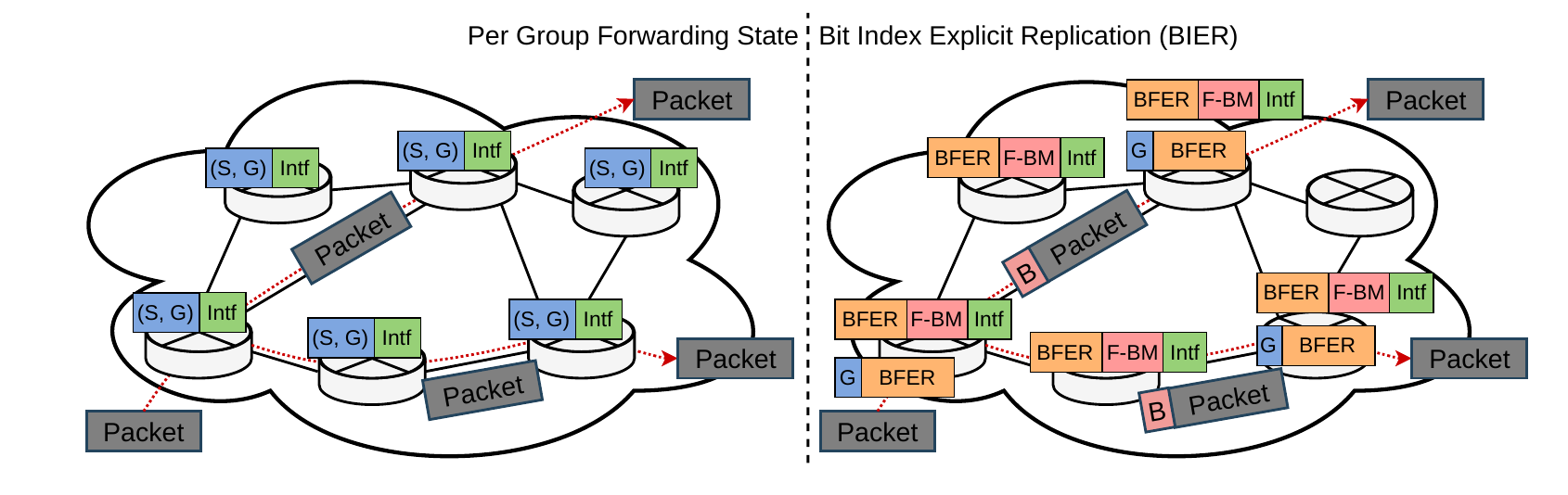}
    \caption{Comparison of different multicast forwarding designs. \textbf{Left:} stateful multicast requires routers to maintain per-group $(S, G)$ forwarding state. \textbf{Right:} Bit Index Explicit Replication (BIER) encapsulates packets with a bit string that identifies egress routers. Core routers forward using a Bit Index Forwarding Table (BIFT) indexed by BFER-id. Each entry specifies a set of outgoing interfaces, each paired with a forwarding bit mask (F-BM) that filters the packet’s bitstring to the destinations reachable via that interface.}
    \label{fig:multicast}
\end{figure*}

\subsection{Setting Forwarding States, Not Routes}

IP multicast introduces a group address: a truly logical identifier that names a set of receivers without specifying their location within the network topology. The delivery of a packet to a multicast group involves mapping this logical identifier to a topological delivery structure, including the set of paths and exit points the packet must traverse to reach all current group members.

It is important to note that multicast routing protocols such as DVMRP~\cite{rfc1075}, MOSPF~\cite{rfc1584}, and PIM~\cite{rfc7761} do not compute independent routing topologies in the same sense as unicast routing protocols. 
Instead, they rely on the underlying unicast routing information to construct multicast distribution trees and install per-group \emph{forwarding state} in routers along those trees.
These forwarding states map $(S,G)$ pairs—source and multicast group identifiers—to sets of outgoing interfaces, thereby associating a logical flow identifier with the underlying network topology.
Each router maintains this state as long as receivers remain in the group, with state creation and removal driven by protocol signaling (\eg join/prune messages) and refreshed periodically.

\subsection{Scaling with Application Dynamics}

The amount and dynamics of multicast forwarding state are largely driven by application-level factors, including the number of active multicast groups, the number of sources per group, group size, and changes in group membership. 
These parameters originate from application behavior rather than network topology, although the resulting forwarding state is instantiated along topology-constrained delivery trees.  
As a consequence, multicast forwarding state offers little opportunity for aggregation compared to unicast routing, where provider-based addressing enables topology-driven aggregation. In source-specific models, each $(S,G)$ pair may require distinct forwarding states at routers along the distribution tree, and even when trees share path segments, their forwarding entries generally cannot be aggregated across groups.  
Thus, the scalability of multicast routing is strongly influenced by application dynamics, which are outside the control of network operators.

Unsurprisingly, multicast deployments in practice have largely been confined to controlled and well-managed environments (\eg enterprise networks and IPTV systems), where forwarding state can be bounded, and application behavior is predictable.
Unlike the unicast case, where hardware and algorithm advances provided a viable incremental path, multicast has no equivalent brute-force escape.
The forwarding state grows with the dynamic behavior of the application, and there is no hardware analog that scales with the application's behavior rather than the network's size.
Deployments have responded by limiting the multicast scope through operational constraints rather than architectural solutions.

\subsection{BIER: Map-and-Encap for Multicast}

Bit Index Explicit Replication (BIER), first proposed by Greg Shepherd at an IETF BOF in November~2014~\cite{shepherd14} and standardized as RFC~8279~\cite{rfc8279} in~2017, provides the multicast equivalent of map-and-encap: an explicit mapping from logical group identifiers to topological egress routers, with the mapping result carried in the packet.

As shown in Figure~\ref{fig:multicast}, when a multicast packet enters a BIER domain, the Bit-Forwarding Ingress Router (BFIR) determines the set of Bit-Forwarding Egress Routers (BFERs) that should receive the packet. This information is derived from multicast control-plane mechanisms (\eg BGP-based distribution) and the unicast routing topology used to reach each BFER, as learned via the IGP. The BFIR then encodes the egress set as a bitstring in the BIER header, where each bit position identifies a BFER within the domain.

This bitstring can be viewed as carrying the delivery specification in the packet: it encodes the set of egress points to which the packet should be delivered. Unlike unicast map-and-encap approaches, which map to a single locator, multicast requires mapping to a set of receivers; BIER achieves this via compact bitstring encoding rather than per-flow state in the core.

Core Bit-Forwarding Routers (BFRs) forward packets using a Bit Index Forwarding Table (BIFT), which is derived from the unicast routing topology. Each BIFT entry maps a bit position (\ie a BFER) to a next hop and outgoing interface, with multiple bits potentially mapping to the same interface. Upon receiving a packet, a BFR examines the bitstring, partitions it by next hop, and forwards one copy per outgoing interface, each carrying only the subset of bits (F-BM) relevant to downstream BFERs. This process ensures efficient replication while delivering exactly one copy to each BFER, without maintaining per-$(S,G)$ state in the core.

Importantly, the BIFT reflects only the network topology and the placement of BFERs, not the set of multicast groups or application-level dynamics. Adding new multicast groups, sources, or receivers does not change the BIFT, provided they map to existing BFERs. 
The mapping from logical multicast membership to topological delivery is thus performed at the ingress, while the core forwarding state remains topology-driven and independent of per-group dynamics.

%

\section{The Scaling Argument}
\label{sec:scaling}
The contrast between stateful multicast and BIER is most clearly
understood through the lens of what each design's resource requirements scale with.
This is the appropriate metric for evaluating any routing architecture: not absolute size but fundamental variables that drive growth.

\subsection{What Drives Forwarding State Growth}

\noindent\textbf{Unicast DFZ forwarding state (FIB)} grows with the number of
unaggregated destination prefixes in the global routing table.
This is driven primarily by the multihoming of the end-site and the portability of the provider, which are administrative choices outside the control of any single network operator.
Hardware advances have managed this growth, but the underlying driver
remains outside of operator control.

\noindent\textbf{Stateful multicast forwarding state} 
grows with application dynamics: the number of active $(S,G)$ pairs, the number of sources
per group, the number of members per group, and the frequency of
membership changes.
These are properties of applications using the network.
They are not under the network operator's control and have no
principled upper bound in an open environment.
The network operator cannot engineer the behavior of the application as it
can engineer the topology of the network.

\noindent\textbf{BIER forwarding state (BIFT)} grows with the number of BIER-capable egress routers in the domain.
The operator determines the scope of a BIER domain, the set of BFRs/BFERs it includes, and how to partition large deployments into multiple bitstring sets (via Set Identifiers or SIs) when the number of BFERs exceeds the bitstring length.
RFC~8279 requires a minimum bitstring length of 256~bits that can
accommodate up to 256~BFERs per SI; larger domains may use multiple SIs.
The forwarding state is bounded by a quantity controlled by the network providers.

\subsection{A General Principle for Evaluating Routing Scalability}
The deeper point is that the network topology is an \emph{engineered} quantity.
ISPs design and operate their networks; they decide how many routers to deploy, how they are interconnected, and how the BIER domain is structured.
The forwarding state required by BIER is therefore bounded by a variable that is under the operator's control and that grows predictably with deliberate network expansion.
In contrast, the number of multicast groups active in a network at any given time depends on the applications deployed, the number of users, and the usage patterns.
The application dynamics is outside of the operator controls in an open environment.  We summarize the above as the following observation:
\begin{quote}
\noindent\textbf{Observation 2: the resource requirements of a routing design should scale with the network topology --- a quantity that can be engineered and controlled --- rather than with application dynamics, which are outside the network's control. 
Designs that violate this principle accumulate architectural debt that operational workarounds can only defer, not eliminate.}
\end{quote}



Routing architecture is fundamentally concerned about where to place
functional complexity to minimize overall system complexity.
Map-and-encap places the mapping function at the network boundary,
where it can be engineered and managed, while simplifying the core,
where states can and should scale with the network rather than with its users.

\section{The Mapping System}
\label{sec:mapping}
Map-and-encap requires a mapping system that binds end-user identifiers to topology identifiers.
The design of this mapping system is just as important as the forwarding architecture itself.

For unicast routing, the early map-and-encap proposal in RFC~1955~\cite{rfc1955} suggested using DNS to map end-user identifiers to network-layer locators.
Although this approach can, in principle, be implemented transparently to applications, it relies on DNS to distribute relatively dynamic locator information and requires host or edge mechanisms to perform mapping and encapsulation.

To avoid exposing this complexity to hosts, LISP introduced a dedicated mapping infrastructure (DDT or ALT), using a pull-based, on-demand resolution model. While this decouples mapping from DNS and preserves application transparency, it introduces a new challenge: the DDT tree requires global administrative coordination at the root, with no agreed-upon global authority, leaving the mapping system without a governance model at the Internet scale.
Alternative approaches such as ILNP~\cite{rfc6740}, which embeds the Identifier/Locator split directly into the IPv6 address structure, avoid this coordination problem but require modifications to the host networking stack, limiting the deployability through a different 
barrier.

ILNP~\cite{rfc6740} takes a different approach by embedding the Identifier/Locator split into the IPv6 address structure and using DNS to distribute locator information.
Although architecturally clean, ILNP requires modifications to host networking stacks and APIs, which has significantly limited its deployability.

In practice, the most successful realization of map-and-encap for unicast is MPLS within an AS, where iBGP provides the mapping between destination prefixes and egress routers.
The mapping problem is tractable here precisely because the scope is bounded: an AS belongs to a specific administrative authority and has engineered boundaries, and egress routers are well-defined topological entities.

Across AS boundaries, inter-domain routing relies on BGP, which distributes reachability information at prefix granularity but does not provide an abstraction for mapping a prefix to a bounded set of egress forwarding points. While BGP can associate a prefix with a set of reachable ASes via AS paths, this information is used for path selection and policy enforcement rather than as a forwarding primitive. Data-plane forwarding remains based on next-hop resolution to specific routers, not to AS-level entities. As a result, map-and-encap approaches are difficult to realize across AS boundaries, since there is no universally agreed-upon topological abstraction at the inter-domain level comparable to an egress router within a single domain.

For multicast routing, BIER applies the same idea to multicast as MPLS to unicast within an AS: IGP distributes bit-index assignments for each BFER, and iBGP (via MVPN extensions~\cite{rfc8556}) carries the mapping between multicast group membership and egress BFERs~\cite{draft-ietf-bier-idr-extensions}.
The bounded scope of a single AS makes this tractable.
\begin{quote}
\noindent\textbf{Observation 3: map-and-encap is the recurring 
architectural answer to routing scalability. MPLS within an AS 
realizes it for unicast; BIER within an AS realizes it for 
multicast. Each was developed independently and in a different 
problem context, yet both arrived at the same principle: resolve 
logical identifiers to topological locators at the domain 
boundary, carry the result in the packet, and let the core 
forward on topology alone.}
\end{quote}

Inter-domain BIER, on the other hand, remains an open 
challenge. Proposed approaches include domain stitching via 
re-encapsulation at domain boundaries~\cite{draft-ietf-bier-multicast-as-a-service}, 
extensions across domains (e.g., BIERv6)~\cite{draft-geng-bier-ipv6-inter-domain}, 
and the use of multicast signaling protocols such as PIM at 
the edges~\cite{draft-ietf-bier-pim-signaling}. These 
approaches tend to reintroduce per-group or per-$(S,G)$ state 
at the domain boundaries or require additional coordination between 
domains.

The reason lies in what an AS actually is.
An AS is primarily an administrative and policy boundary, rather than a 
well-defined topological unit.
Consequently, there is no universally defined notion of an ``exit AS'' that can serve as a stable mapping target analogous to an egress router within a 
domain.
Without such an abstraction, mapping systems cannot fully aggregate forwarding state at the inter-domain level, and the scalability benefits observed within a single domain do not extend across AS boundaries.
Both BGP and all proposed inter-domain BIER designs fall back to finer-grained  state at domain boundaries for the same reason.
\begin{quote}
\noindent\textbf{Observation 4: inter-domain routing lacks a topological abstraction analogous to an egress router within a single domain. Consequently, both unicast (via BGP) and multicast (in proposed inter-domain BIER designs) must fall back to finer-grained state at domain boundaries.}
\end{quote}

\vspace{-0.4cm}
\section{Conclusion}
\label{sec:conclusion}

The Internet's routing scalability problems, in both unicast and multicast, share a common architectural root: the conflation of end-user identifier with topological locator in a single IP address space. Rekhter's Law captured this precisely, but the Internet chose neither cleanly, and the resulting tension persists in the challenges documented in this paper.

The four observations this paper makes point in a common direction.
Map-and-encap is the right architectural answer -- the community has arrived at it repeatedly and independently, for unicast and multicast alike.
Designs that scale with application dynamics accumulate architectural debt that operational workarounds can only defer, not eliminate. 
Designs that scale with topology --- a quantity operators engineer and control --- are fundamentally sound. 
BIER demonstrates that this principle can be realized cleanly for multicast, three decades after the community first articulated it for unicast. 
Yet no deployed solution exists at the global scale, for the same underlying reason in both cases: inter-domain routing lacks a topological abstraction analogous to an egress router within a single domain, and building one requires global coordination that no single operator can unilaterally drive.
LISP's DDT is the clearest illustration of this barrier.

The remaining challenge is to extend it beyond the boundaries of a single provider domain.
The experience with INLP and LISP suggest that a new global mapping system cannot be retrofitted into a deployed routing infrastructure; it needs to be designed in as a system component from the start. 
For any future inter-domain architecture that takes routing scalability seriously, this should be among the foundational considerations.



\bibliographystyle{ACM-Reference-Format}
\bibliography{refs}

@inproceedings{clark88,
  author    = {David D. Clark},
  title     = {The Design Philosophy of the {DARPA} Internet Protocols},
  booktitle = {Proceedings of the ACM SIGCOMM Conference},
  pages     = {106--114},
  month     = aug,
  year      = {1988}
}

@techreport{rfc1955,
  author      = {Robert Hinden},
  title       = {New Scheme for Internet Routing and Addressing ({ENCAPS})
                 for {IPNG}},
  institution = {Internet Engineering Task Force},
  type        = {RFC},
  number      = {1955},
  month       = jun,
  year        = {1996},
  url         = {https://www.rfc-editor.org/rfc/rfc1955}
}

@techreport{rfc4984,
  author      = {David Meyer and Lixia Zhang and Kevin Fall},
  title       = {Report from the {IAB} Workshop on Routing and Addressing},
  institution = {Internet Engineering Task Force},
  type        = {RFC},
  number      = {4984},
  month       = sep,
  year        = {2007},
  url         = {https://www.rfc-editor.org/rfc/rfc4984}
}

@techreport{rfc7761,
  author      = {Bill Fenner and Mark Handley and Hugh Holbrook and
                 Isidor Kouvelas and Rishabh Parekh and Zhaohui Zhang and
                 Lianshu Zheng},
  title       = {Protocol Independent Multicast --- Sparse Mode ({PIM-SM}):
                 Protocol Specification (Revised)},
  institution = {Internet Engineering Task Force},
  type        = {RFC},
  number      = {7761},
  month       = mar,
  year        = {2016},
  url         = {https://www.rfc-editor.org/rfc/rfc7761}
}

@techreport{rfc8279,
  author      = {IJsbrand Jan Wijnands and Eric Rosen and Andrew Dolganow and
                 Tony Przygienda and Sam Aldrin},
  title       = {Multicast Using Bit Index Explicit Replication ({BIER})},
  institution = {Internet Engineering Task Force},
  type        = {RFC},
  number      = {8279},
  month       = nov,
  year        = {2017},
  url         = {https://www.rfc-editor.org/rfc/rfc8279}
}

@techreport{rfc9300,
  author      = {Dino Farinacci and Vince Fuller and Dave Meyer and
                 Darrel Lewis and Alberto Cabellos},
  title       = {The Locator/{ID} Separation Protocol ({LISP})},
  institution = {Internet Engineering Task Force},
  type        = {RFC},
  number      = {9300},
  month       = oct,
  year        = {2022},
  url         = {https://www.rfc-editor.org/rfc/rfc9300}
}

@techreport{draft-ietf-bier-idr-extensions,
  author       = {Xu, X. and Przygienda, T. and Gulko, A. and
                  Satapati, S. and Wijnands, I.},
  title        = {{BGP Extensions for BIER}},
  institution  = {IETF},
  type         = {Internet-Draft},
  number       = {draft-ietf-bier-idr-extensions-19},
  year         = {2024},
  month        = dec,
  note         = {Work in Progress},
  url          = {https://datatracker.ietf.org/doc/draft-ietf-bier-idr-extensions/}
}

@techreport{draft-ietf-bier-multicast-as-a-service,
  author       = {Zhang, Z. and Rosen, E. and Awduche, D. and
                  Shepherd, G.},
  title        = {{Multicast/BIER As A Service}},
  institution  = {IETF},
  type         = {Internet-Draft},
  number       = {draft-ietf-bier-multicast-as-a-service-03},
  year         = {2022},
  month        = jul,
  note         = {Work in Progress},
  url          = {https://datatracker.ietf.org/doc/draft-ietf-bier-multicast-as-a-service/}
}

@techreport{draft-geng-bier-ipv6-inter-domain,
  author       = {Geng, L. and Xie, J. and McBride, M. and
                  Yan, G. and Geng, X.},
  title        = {{Inter-Domain Multicast Deployment using BIERv6}},
  institution  = {IETF},
  type         = {Internet-Draft},
  number       = {draft-geng-bier-ipv6-inter-domain-02},
  year         = {2020},
  month        = oct,
  note         = {Expired},
  url          = {https://datatracker.ietf.org/doc/draft-geng-bier-ipv6-inter-domain/}
}

@techreport{draft-ietf-bier-pim-signaling,
  author       = {Bidgoli, H. and Venaas, S. and Mishra, G. and
                  Zhang, Z. and McBride, M.},
  title        = {{PIM Signaling Through BIER Core}},
  institution  = {IETF},
  type         = {Internet-Draft},
  number       = {draft-ietf-bier-pim-signaling-13},
  year         = {2025},
  month        = mar,
  note         = {Work in Progress},
  url          = {https://datatracker.ietf.org/doc/draft-ietf-bier-pim-signaling/}
}

@misc{shepherd14,
  author       = {Greg Shepherd},
  title        = {Bit Index Explicit Replication ({BIER})},
  howpublished = {Presented at IETF MBONED BOF, IETF~91, Honolulu, HI},
  month        = nov,
  year         = {2014},
  url          = {https://www.ietf.org/proceedings/92/slides/slides-92-mboned-5.pdf}
}

@techreport{rfc6740,
    author = {Randall J. Atkinson and Saleem Bhatti},
    title = {{Identifier-Locator Network Protocol (ILNP) Architectural Description}},
    howpublished = {Internet Requests for Comments},
    type = {RFC},
    number = {6740},
    year = {2012},
    month = {November},
    publisher = {RFC Editor},
    institution = {RFC Editor},
    url = {https://datatracker.ietf.org/doc/rfc6740/}
}

@techreport{rfc8556,
    author = {Eric C. Rosen and Mahesh Sivakumar and Sam Aldrin and Andrew Dolganow and Tony Przygienda},
    title = {{Multicast VPN Using Bit Index Explicit Replication (BIER)}},
    howpublished = {Internet Requests for Comments},
    type = {RFC},
    number = {8556},
    year = {2019},
    month = {April},
    publisher = {RFC Editor},
    institution = {RFC Editor},
    url = {https://www.rfc-editor.org/info/rfc8556}
}

@misc{fuller2007,
  author       = {Vince Fuller},
  title        = {The Future of Routing Scalability},
  howpublished = {APRICOT 2007 Conference},
  year         = {2007},
  note         = {Presentation slides},
  url          = {https://www.apricot.net/apricot2007/presentation/apia-future-routing/apia-future-routing-vince-fuller.pdf}
}

@techreport{rfc3031,
  author       = {E. Rosen and A. Viswanathan and R. Callon},
  title        = {Multiprotocol Label Switching Architecture},
  institution  = {IETF},
  type         = {RFC},
  number       = {3031},
  year         = {2001},
  url          = {https://datatracker.ietf.org/doc/html/rfc3031}
}

@techreport{rfc6830,
  author       = {D. Farinacci and V. Fuller and D. Meyer and D. Lewis},
  title        = {The Locator/ID Separation Protocol (LISP)},
  institution  = {IETF},
  type         = {RFC},
  number       = {6830},
  year         = {2013},
  url          = {https://datatracker.ietf.org/doc/html/rfc6830}
}

@techreport{rfc6833,
  author       = {V. Fuller and D. Farinacci},
  title        = {Locator/ID Separation Protocol (LISP) Map-Server Interface},
  institution  = {IETF},
  type         = {RFC},
  number       = {6833},
  year         = {2013},
  url          = {https://datatracker.ietf.org/doc/html/rfc6833}
}

@techreport{rfc9301,
  author       = {D. Farinacci and F. Maino and V. Fuller and A. Cabellos},
  title        = {Locator/ID Separation Protocol (LISP) Control Plane},
  institution  = {IETF},
  type         = {RFC},
  number       = {9301},
  year         = {2022},
  url          = {https://datatracker.ietf.org/doc/html/rfc9301}
}

@techreport{rfc9303,
  author       = {F. Maino and V. Ermagan and A. Cabellos and D. Saucez},
  title        = {Locator/ID Separation Protocol Security (LISP-SEC)},
  institution  = {IETF},
  type         = {RFC},
  number       = {9303},
  year         = {2022},
  url          = {https://datatracker.ietf.org/doc/html/rfc9303}
}

@techreport{rfc1584,
  author       = {J. Moy},
  title        = {Multicast Extensions to OSPF},
  institution  = {IETF},
  type         = {RFC},
  number       = {1584},
  year         = {1994},
  url          = {https://datatracker.ietf.org/doc/html/rfc1584}
}

@techreport{rfc1075,
  author       = {D. Waitzman and C. Partridge and S. Deering},
  title        = {Distance Vector Multicast Routing Protocol},
  institution  = {IETF},
  type         = {RFC},
  number       = {1075},
  year         = {1988},
  url          = {https://datatracker.ietf.org/doc/html/rfc1075}
}

\end{document}